\documentclass[usenatbib]{mnras}

\usepackage[T1]{fontenc}
\usepackage{ae,aecompl,color,dblfloatfix}

\usepackage{graphicx}      
\usepackage{epstopdf}
\usepackage{amsmath}    
\usepackage{amssymb}    
\usepackage{booktabs}
\usepackage{bm}

\definecolor{red}{rgb}{0.8,0,0}
\definecolor{violet}{rgb}{0.4,0,0.4}
\definecolor{green}{rgb}{0,0.5,0.0}
\definecolor{navy}{rgb}{0.0,0.0,0.6}
\definecolor{orange}{rgb}{0.8,0.2,0.0}

\usepackage[normalem]{ulem}  

\def\apj{ApJ \ }%
\def\apjl{ApJ Lett.}%
\def\apss{Ap\&SS}%
\def\aap{A\&A}%
\def\prc{Phys. Rev. C\ }%
\def\prd{Phys. Rev. D \ }%
\def\nat{Nature}%
\title[Cooling of hypernuclear stars]{                                                                                                   
Cooling of hypernuclear compact stars:}

\author[A. R. Raduta et al. ]{
Adriana R. Raduta$^{1}$\thanks{E-mail: araduta@nipne.ro}
Armen Sedrakian $^{2}$\thanks{sedrakian@fias.uni-frankfurt.de}
and Fridolin Weber $^{3,4}$\thanks{fweber@sdsu.edu}
\\
$^{1}$National Institute for Physics and Nuclear Engineering,  RO-077125 Bucharest, Romania\\
$^{2}$Frankfurt Institute for Advanced Studies, D-60438 Frankfurt-Main, Germany\\
$^{3}$Department of Physics, San Diego State University, 5500
Campanile Drive, San Diego, CA 92182, USA \\
$^{4}$Center for Astrophysics and Space Sciences, University of California at San
Diego, La Jolla, CA 92093, USA 
}

\begin{document}
\label{firstpage}
\pagerange{\pageref{firstpage}--\pageref{lastpage}}
\maketitle

\begin{abstract}
  We study the thermal evolution of hypernuclear compact stars
  constructed from covariant density functional theory of hypernuclear
  matter and parameterizations which produce sequences of stars
  containing two-solar-mass objects. For the input in the simulations,
  we solve the Bardeen-Cooper-Schrieffer gap equations in the
  hyperonic sector and obtain the gaps in the spectra of $\Lambda$,
  $\Xi^0$ and $\Xi^-$ hyperons. For the models with masses
  $M/M_{\odot} \ge 1.5$ the neutrino cooling is dominated by hyperonic
  direct Urca processes in general. In the low-mass stars the
  $(\Lambda p)$ plus leptons channel is the dominant direct Urca
  process, whereas for more massive stars the purely hyperonic
  channels $(\Sigma^-\Lambda)$ and $(\Xi^-\Lambda)$ are dominant.
  Hyperonic pairing strongly suppresses the processes on $\Xi^-$s and
  to a lesser degree on $\Lambda$s. We find that intermediate-mass
  $1.5 \le M/M_{\odot} \le 1.8$ models have surface temperatures which
  lie within the range inferred from thermally emitting neutron stars,
  if the hyperonic pairing is taken into account. Most massive models
  with $M/M_{\odot} \simeq 2$ may cool very fast via the direct Urca
  process through the $(\Lambda p)$ channel because they develop inner
  cores where the $S$-wave pairing of $\Lambda$s and proton is absent.
\end{abstract}


\section{Introduction}
The observations of several white dwarf--pulsar binaries with pulsar
masses close to two solar masses
\citep{2010Natur.467.1081D,2013Sci...340..448A,Fonseca2016,2017MNRAS.465.1711B}
have spurred intensive research on the problem of hyperonization of
dense matter in compact stars.  The key issue is the construction of
models of compact stars containing hypernuclear matter in their cores,
which accommodate the two-solar-mass compact stars mentioned just
above.  The core compositions of compact stars are computed for a given
nuclear equation of state (hereafter EoS). Depending on mass, 
nuclear densities in excess of several times the density of normal
nuclear matter are encountered.  So far, the bulk of the research on
such stars has been directed towards the integral parameters of
non-rotating and rotating compact
stars~\citep{Weissenborn2012b,Weissenborn2012a,Bonanno2012,Bednarek2012,Long2012,2013PhRvC..87e5806C,Miyatsu_PRC2013,2014PhLB..734..383V,Gusakov_MNRAS2014,2015PhLB..748..369M,2015ApJ...808....8G,2015JPhG...42g5202O,Fortin_PRC2016,Tolos2016,2017arXiv170602913M}.

The purpose of this work is to advance these studies by addressing the
problem of their thermal evolution.  If the late-time heating
processes are ignored then the problem reduces to the neutrino cooling
from stellar interior during the time-span $t\le 10^5$~yr after the
star's birth, which is followed by asymptotic photon cooling from its
surface. The onset of hyperons in dense matter gives rise to an array
of new processes involving weak decays of hyperons (and their
inverse), such as direct~\citep{1992ApJ...390L..77P} and modified Urca
processes~\citep{1987ApJ...316..691M,2016Ap&SS.361..267K}. The direct
Urca (hereafter dUrca) processes strongly enhance the neutrino
luminosity of the star, potentially cooling it very
rapidly~\citep{Boguta1981,Lattimer1991}. The required
threshold densities of hyperons for these processes to operate are very
low, of the order of several percent of the density.  Therefore these
processes become operative at densities slightly above those where
hyperons first become energetically favourable in compact star matter,
provided that all involved species are present.

Early models of compact star cooling with hyperon admixtures were
studied by \cite{Haensel_1994} in the isothermal approximation for
non-superfluid hyperons and by \cite{Schaab_1998} using
  a non-isothermal code which accounted for $\Lambda$
hyperon pairing. A more recent study by \cite{Tsuruta_2009} included
also the pairing of $\Sigma^-$ hyperons along with other additional physics, such as three-body forces in
the EoS. Since the onset of hyperons
softens the EoS drastically and lowers the
maximum mass that can be supported by the EoS, the cooling models
mentioned above were built for stars with relatively low masses (e.g.,
in \cite{Tsuruta_2009} the sequences are restricted to $M/M_{\odot}\le
1.7$), which are contrary to present-day measurements.
On the other hand, the relativistic density functionals (DF) constructed
recently  
provide models of hypernuclear
stars which satisfy the presently known constraints from laboratory
physics and astrophysics of compact stars. It is,
therefore, the purpose of this work to study the
cooling of compact stars whose mass range is in agreement with
observed data by employing models for the nuclear EoS which account
for hyperonic degrees of freedom, and to unveil the new
characteristics which hyperonization has on the cooling of compact
stars.

This work is structured as follows. In Sec.~\ref{sec:pairing} we first
set the stage by describing the covariant DFs on which
the EoS used in this work are based. We then go on to solve the
Bardeen-Cooper-Schrieffer (BCS) equations in the hyperonic sector to
obtain the gaps and critical temperatures for hyperons
(i.e, $\Lambda$ and $\Xi^{-,0}$) interacting via attractive forces.  
Section~\ref{sec:nu_processes} is
devoted to the discussion of neutrino processes introduced by the
hyperonic component; we list the main direct Urca processes on
hyperons, as well provide updated rates of the
pair-breaking processes on hyperons which account for suppression of
the vector current contributions.  Section~\ref{sec:therm_ev}
describes the results of our simulations of the thermal
evolution of hypernuclear compact stars for three models from our
EoS collection.  Our conclusions and a concise summary
can be found in Sec.~\ref{sec:conclusions}.

\section{Equation of state and pairing}
\label{sec:pairing}

\subsection{Density functionals for hypernuclear matter}
\label{subsec:DF}

Hyperons in dense nuclear matter have been studied using a number of
methods, ranging from non-relativistic potential based many-body
models to Lagrangian based relativistic density functional (DF)
methods~\citep{weber_book,2007PrPNP..58..168S}. The parameters of DFs
are fixed by the nuclear phenomenology of hypernuclear matter, nuclear
collisions, and compact stars. Non-relativistic potential 
models \citep{Balberg_1997,Baldo2000,Burgio2011}
fail to produce heavy  enough neutron stars (NS) and/or are 
incompatible with most recent experimental hypernuclear data.

The relativistic DF formalism provides a consistent theoretical
framework, which can be used to extrapolate the nuclear EoS to very
high densities.  In this work we use a set of representative DFs based
on the density-dependent parametrization of nucleonic DFs,
specifically the DDME2 parametrization~\citep{2005PhRvC..71b4312L},
and DFs which have constant coupling constants but include non-linear
mesonic contributions instead, such as NL3~\citep{PhysRevC.55.540} and
GM1~\citep{1991PhRvL..67.2414G}.  The extensions of the DDME2 model to
the hypernuclear sector have been carried out in several works
\citep{2013PhRvC..87e5806C,2014PhLB..734..383V,Fortin_PRC2016,Spinella-PhD} 
and we shall use the parametrization
of \cite{Fortin_PRC2016} and \cite{Spinella-PhD} below; we shall adopt
the parameter set of \cite{Gusakov_MNRAS2014} for GM1A model; for NL3
model we shall employ the parameter set NL3(b) of
\cite{Miyatsu_PRC2013}, which is identical to NL3Yss of
\cite{Fortin_PRC2016} and very similar to the hyperonic NL3 model of
\cite{Wang_PRC2010}.

\begin{table}
\begin{tabular}{lcccccc}
\hline 
Model& $n_{\rm s}$       & $E_{\rm s}$ & $K$  & $J$  & $L$   &$K_{\rm  sym}$ 
                                                                         \\
     & [fm$^{-3}$]      & [MeV]     & [MeV] & [MeV]& [MeV] & [MeV]      \\
\hline 
NL3  &  0.149          & -16.2     & 271.6 & 37.4 & 118.9 & 101.6       \\
GM1A & 0.154           & -16.3     & 300.7 & 32.5 & 94.4  & 18.1         \\
DDME2& 0.152           & -16.1     & 250.9 & 32.3 & 51.2  & -87.1      \\
SWL  & 0.150           & -16.0     & 260.0 & 31.0 & 55.0  & n.a.        \\
\hline                                                                                                                                    
\end{tabular}
\caption{Key nuclear properties of the relativistic DF models 
  considered in this work. Listed are the energy per nucleon ($E_s$) and
  compression modulus ($K$) at the saturation density of symmetric nuclear matter ($n_s$)
  together with the symmetry energy ($J$), 
  slope ($L$) and curvature ($K_{\rm sym}$) of the symmetry energy 
  at $n_s$.}
\label{tab:models}
\end{table}
\begin{table*}
\begin{tabular}{ll c cc cc cc cc cc c cc}
\hline 
Model &  mesons & flavor & $n_{\rm max}$& $M_{\rm max}^{ Y}$ & $Y_1$ & $n_{Y_1}$ & $M_{Y_1}$
                                                 & $Y_2$ & $n_{Y_2}$ & $M_{Y_2}$
                                                 & $Y_3$ & $n_{ Y_3}$ & $M_{Y_3}$  & $n_{\rm DU}$ &  $M_{\rm DU}$ \\
      &  & sym. & [fm$^{-3}$]& [$M_{\odot}$]      &       & [fm$^{-3}$]  & [$M_{\odot}$]     &
      & [fm$^{-3}$] &  [$M_{\odot}$]      &   & [fm$^{-3}$] &
                                                              [$M_{\odot}$]  & [fm$^{-3}$] &  [$M_{\odot}$]
\\
\hline 
NL3 & $\sigma$,$\sigma^*$,$\omega$,$\phi$,$\rho$ & SU(6) & 0.77 & 2.07 & $\Lambda$  & 0.28 & 1.47 
& $\Xi^-$ & 0.33 & 1.73 
& $\Xi^0$ & 0.57 & 2.02 & 0.21       & 0.85\\
GM1A & $\sigma$,$\omega$,$\phi$,$\rho$ & SU(6) & 0.92 & 1.994 & $\Lambda$ & 0.35 & 1.49 
& $\Xi^-$ & 0.41 & 1.67 
& - & - & - & 0.28       & 1.10\\
DDME2 &  $\sigma$,$\omega$,$\phi$,$\rho$  & SU(6)          & 0.93 & 2.12 & $\Lambda$ & 0.34 & 1.39 
& $\Xi^-$ & 0.37 & 1.54 & $\Sigma^-$ & 0.39 & 1.60     & -          & - \\
SWL    & $\sigma$,$\omega$,$\rho$   & SU(3)               & 0.97 & 2.003& $\Lambda$ & 0.41 & 1.51 
& $\Xi^-$ & 0.45 & 1.65 & $\Xi^0$ & 0.90 & 2.00  & 0.90       & 2.00\\
\hline                                                            
\end{tabular}
\caption{Astrophysical characteristics of the
    relativistic DF EoS models (with hyperons) used in this work:
    $n_{\rm max}$ shows the central densities of the maximum-mass
    ($M_{\rm max}^Y$) hyperonic star of each stellar sequence,
    $n_{Y_i}$ shows the threshold densities at which hyperons of type
    $Y_i$ are produced, and $M_{Y_i}$ lists the mass of the hyperonic
    star for that density.  The last two entries show the baryon
    number density ($n_{DU}$) beyond which the nucleonic dUrca process
    is allowed in purely nucleonic NS matter and the mass
    ($M_{DU}$) of the associated compact star.}
\label{tab:rmfY}
\end{table*}

In Table~\ref{tab:models} we list the nuclear parameters of these
models. Note that, in the NL3 model, the saturation values of the
symmetry energy $J$ and its slope $L$ are outside of the preferred
ranges~\citep{Tsang_2012,Lattimer_2014}.  Nevertheless, we keep this
model in our collection for the sake of illustration.

Table \ref{tab:rmfY} displays the properties of the hyperonic stars
computed for our collection of EoSs.  For each DF the second columns
lists mesonic fields whereas the third column specifies the underlying
flavor symmetry group.  The couplings of the $\sigma$ scalar meson to
hyperons, $g_{\sigma Y}$, are typically determined from the values
of semi-empirical depths of potential wells for hyperons at rest in
symmetric nuclear matter at saturation density,
\begin{equation} 
U_{Y}^{(N)}(n_s)=-\left( g_{\sigma Y} +g'_{\sigma Y} \rho_s \right) \sigma 
+ \left( g_{\omega Y} +g'_{\omega Y} n_s \right) \omega,
\end{equation}
where $\rho_s$ is the scalar density, the prime denotes the derivative
with respect to the total density; the derivative terms are non-zero
only in the models with density-dependent couplings.  In all
DFs the following values of hyperonic potentials were
used: $U_{\Lambda}^{(N)} \approx -28$ MeV, $U_{\Xi}^{(N)} \approx -18$
MeV and $U_{\Sigma}^{(N)} \approx 30$ MeV \citep{Millener1988}.  In
the NL3  model \citep{Fortin_PRC2016,Miyatsu_PRC2013,Wang_PRC2010}, which
accounts for the hidden strangeness meson $\sigma^*$,
$g_{\sigma^* \Lambda}$ is determined from the value of the
$\Lambda$-potential in $\Lambda$ matter
\begin{equation}
U_{\Lambda}^{(\Lambda)}=-g_{\sigma \Lambda} \sigma- g_{\sigma^* \Lambda} \sigma^*
+g_{\omega \Lambda} \omega+g_{\phi \Lambda} \phi,
\end{equation}
assuming $U_{\Lambda}^{(\Lambda)}(n_s) \approx -5$ MeV
\citep{Takahashi2001}.  The couplings with the other hyperons are
obtained from symmetry arguments,
$g_{\sigma^* \Sigma}=g_{\sigma^* \Lambda}$ and
$U_{\Xi}^{(\Xi)} \approx 2 U_{\Lambda}^{(\Lambda)}$.  The vector
meson-hyperon couplings constants are expressed in terms of the
couplings to the nucleon and assume certain flavor symmetries.  The
first three DFs in Tables \ref{tab:models} and \ref{tab:rmfY} are
based on the SU(6) and the SWL on the SU(3) flavour symmetry.  In this
last case, the ESC08 model~\citep{Rijken2013} values are employed for
the vector mixing angle $\theta_v=37.50^o$, the vector coupling ratio
$\alpha_V=1$ and the meson singlet to octet coupling ratio
$z=0.79$.

Columns 4-14 of Table \ref{tab:rmfY} list the maximum mass and the
corresponding central baryon number density, the threshold density of
each hyperonic species and the mass of the star associated with that
density.  It is seen that for all models the first two hyperons to appear
are the $\Lambda$ and $\Xi^-$ hyperons; the third type of hyperon to
nucleate could be either $\Xi^0$ or $\Sigma^-$.  The chosen
DFs favour $\Xi^-$ hyperon over the less massive $\Sigma^-$ because of
its potential in nuclear matter is attractive, whereas that of the
$\Sigma^-$ is repulsive~\citep{Millener1988}.  Clearly, if the
threshold density for appearance of any given hyperon is larger than
the central density of the maximum-mass star, that particular hyperon
will not be accounted in our simulations.  Finally, the last two
columns of Table \ref{tab:rmfY} list the baryon number density and NS
mass threshold above which the nucleonic dUrca process operates in a
purely nucleonic NS matter.

\subsection{BCS models of hyperonic pairing}
\label{subsec:pairing}

The attractive component of the nuclear force between the hyperons
will lead to their BCS pairing. Because of
the relatively low density of hyperons, the most attractive partial
wave is the $^1S_0$ channel, which would pair hyperons as spin-singlet
Cooper pairs.  Hyperon-nucleon pairing as well as pairing among
non-identical hyperons, e.g. $\Lambda\Xi^{-,0}$ will be suppressed
because of the difference in their densities and/or the difference in
their effective masses~\citep{Stein2014}.  In fact, the latter
strongly disfavour cross-species pairing among baryons even when their
abundances become equal at some density.

A reliable strategy for computing pairing gaps in nucleonic matter has
been outlined in past studies of relativistic DF models of nucleonic
pairing, where the non-relativistic BCS equation is solved for a given
two-nucleon potential~\citep{1991ZPhyA.339...23K} using
single-particle energies and particle composition computed for the
relativistic DF method. Although there is certain inconsistency in the
methods of treating the background and pairing correlations, this
approach has been validated in computations of finite nuclei within
the relativistic Hartree-Fock-Bogoliubov theory~\citep{LongRing2010}.

We employ this strategy and use for the composition of matter the DF
results from the previous subsection; for the $\Lambda\Lambda$ pairing
interaction we use the configuration space parametrization of ESC00
potentials~\citep{2001NuPhA.691..322R} given by
\cite{Filikhin_NPA2002}; for $\Xi^{-}\Xi^{-}$ and $\Xi^{0}\Xi^{0}$
interaction we use the potential designed by
\cite{2016PhRvC..94b4002G}, which corresponds to Nijmegen Extended
Soft Core ESC08c potential~\citep{Rijken2013}.  Our choice of ESC00
and ESC08c potentials is motivated by the fact that they provide
maximum attraction in the $\Lambda\Lambda$ and, respectively, $\Xi\Xi$
channels.  Consequently, our results provide an upper limit on the
hyperon pairing and thus maximize the role of hyperon pairing on NS
cooling. The $\Sigma \Sigma$ pairing is disregarded because according
to ESC08c potential this interaction channel is repulsive.  The
pairing in the $\Lambda$-channel was studied by \cite{Balberg_PRC1998}
and most recently by \cite{Wang_PRC2010} for a matter composition
determined from relativistic DF theory. The $\Xi$-channel has been
briefly discussed by \cite{Takatsuka_XiPairing}, but the physical
implications of $\Xi$ pairing has remained largely unexplored.

The quantity determining the onset of superfluidity is the energy gap
function $\Delta_{Y}(k)$, obtained by solving the gap equation,
\begin{equation}
  \Delta_{Y} (k)=-\frac{1}{4 \pi^2} \int dk' k'^2 \frac{V_{YY}(k,k') 
\Delta_{Y}(k')}{\sqrt{\left[E_{s.p.}^Y(k')-\mu_Y \right]^2+\Delta_{Y}^2(k')}},
\label{eq:gap}
\end{equation}
where $E_{s.p.}(k)$ is the single-particle energy of hyperon $Y$
with momentum $k$,
\begin{equation}
E^Y_{s.p.}(k)=\sqrt{(\hbar c)^2 k^2+m_Y^{*2}}+g_{\omega Y} 
\omega+g_{\phi Y} \phi+g_{\rho Y} \tau_{3Y} \rho+\Sigma_R,
\label{eq:esp}
\end{equation}
where $\Sigma_R$ represents the rearrangement term entering the models with
density-dependent couplings, $\mu_Y=E^Y_{s.p.}(k_F)$ stands for the
chemical potential and
$m_Y^*=m_Y-g_{\sigma Y} \sigma - g_{\sigma^*Y} \sigma^*$ is the Dirac
effective mass of the species $Y$.  For the pairing interaction in the
$^1S_0$ channel the potential matrix element can be written as
\begin{equation}
  V_{YY}(k,k')=\langle k | V_{YY} | k'\rangle= 4 \pi \int dr r^2 j_0(k r) V_{YY}(r) j_0(k' r),
\label{eq:Vkk}
\end{equation}
where $j_0(k r)=\sin(kr)/(kr)$ is the spherical Bessel function of
order zero and $V_{YY}(r)$ is the $^1S_0$ channel $YY$ interaction
potential in coordinate space.  The gap equation \eqref{eq:gap} was
solved numerically by using as an input Eq.~\eqref{eq:Vkk}, with the
configuration space interactions for the $\Lambda$ and $\Xi$ channels
taken from \cite{Filikhin_NPA2002} and \cite{2016PhRvC..94b4002G},
respectively, and for matter properties computed for the models
introduced in Sec.~\ref{subsec:DF}.  An iterative method for solving
the gap equation was applied with adaptive momentum mesh to account
for rapid variations of the integrand in the vicinity of Fermi
momentum~\citep{2006PhRvC..73c5803S}.

\begin{figure}
\begin{center}
\includegraphics[angle=0, width=0.89\columnwidth]{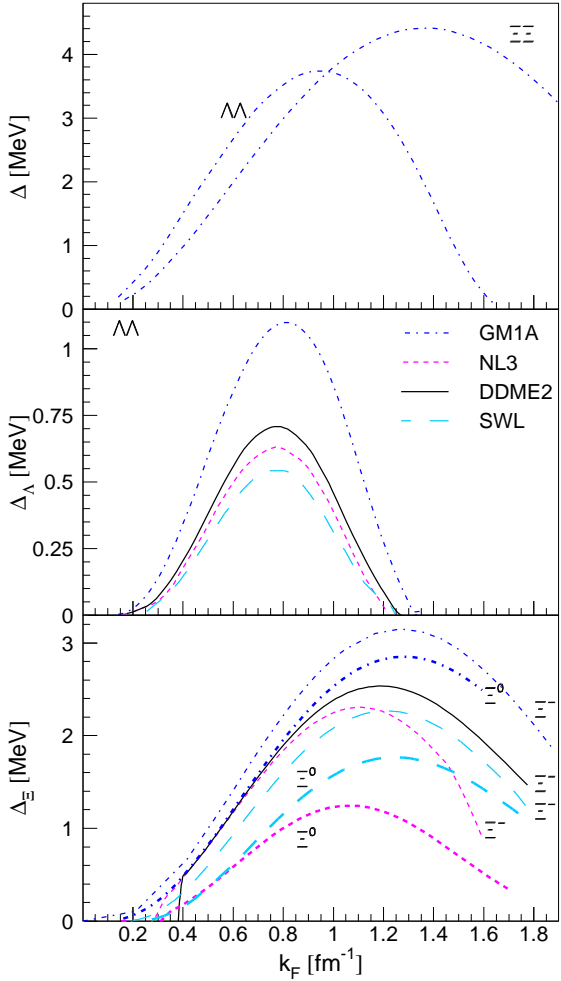}
\end{center}
\caption{Dependence of $^1S_0$ pairing gaps at Fermi energy on the
  Fermi momentum for $\Lambda$ and $\Xi$ hyperons in NS matter based
  on our collection of DF models: (top) gaps obtained with $m^*=m$ and
  chemical potentials taken from GM1A model; (middle) gaps for
  $\Lambda$ hyperons with medium dependent single particle energies
  and composition according to DDME2 (solid), NL3 (dashed), GM1A
  (dash-dotted) and SWL (long-dashed) models; (bottom) same as middle
  panel, but for $\Xi^{-}$ (thin lines) and $\Xi^{0}$ (thick lines)
  hyperons. Note that in the case of DDME2 DF the $\Xi^{0}$ hyperon
  does not nucleate in the density range shown in the figure. }
\label{fig:Gaps_kF}
\end{figure}

Fig.~\ref{fig:Gaps_kF} shows the dependence of the pairing gaps for
$\Lambda$ and $\Xi^{-,0}$ hyperons on their respective Fermi momenta.
When the dispersive effects are neglected (Fig.~\ref{fig:Gaps_kF}, top
panel) the gaps reflect the attraction in the given channel. The
bell-shaped form of these curves results from the increase in the
density of states combined with the decreasing attraction among the
hyperons as their Fermi momenta $k_F$ increase. The reduction of the
hyperon masses by the medium reduces the density of states, and hence
the gap at the Fermi surface.  This effect is more pronounced at
higher densities where the effective masses are substantially smaller
than unity (middle and bottom panels of Fig.~\ref{fig:Gaps_kF}). The
reduction in the $\Lambda\Lambda$ pairing is larger than in the case
of $\Xi^-\Xi^-$ and $\Xi^0\Xi^0$ and reflects the magnitude of the
change in the effective mass. We have computed also pairing gaps using
as a background other DFs, which were recently proposed in the
literature, e.g. GM1'B and TM1C from \cite{Gusakov_MNRAS2014} and
GM1(c), TM1(c) and NL3(c) from \cite{Miyatsu_PRC2013}, all of which
fulfill recent hypernuclear and astrophysical constrains. For these
DFs the pairing gaps are comparable or lower than those shown in
Fig. ~\ref{fig:Gaps_kF}. 

\begin{figure}
\begin{center}
\includegraphics[angle=0, width=0.89\columnwidth]{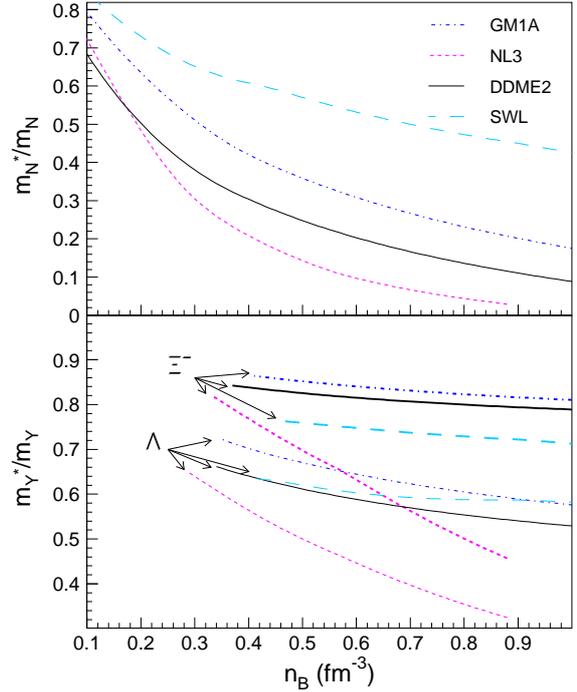}
\end{center}
\caption{Dependence of effective masses of neutrons (top panel) and of
  $\Lambda$ (thin) and $\Xi^-$ (thick) hyperons (bottom panel) on
  baryonic number density, for DF models. The proton and $\Xi^0$
  effective masses are close to those of neutrons and $\Xi^-$ hyperons
  as none of our models accounts for $\delta$-mesons. }
\label{fig:effmesses}
\end{figure}

To illustrate the suppression of the pairing gaps by dispersive
effects quantitatively we show in Fig.~\ref{fig:effmesses} the
dependence of the effective masses of baryons on the baryon density
for the models studied in this work.  The effective masses of baryons
decrease with increasing density due to their interactions with
 scalar mesons. (The NL3 model leads to an unrealistic drop of
the effective mass, but this occurs at densities where no stable
configurations of compact stars exist.)

\begin{figure}
\begin{center}
\includegraphics[angle=0, width=0.89\columnwidth]{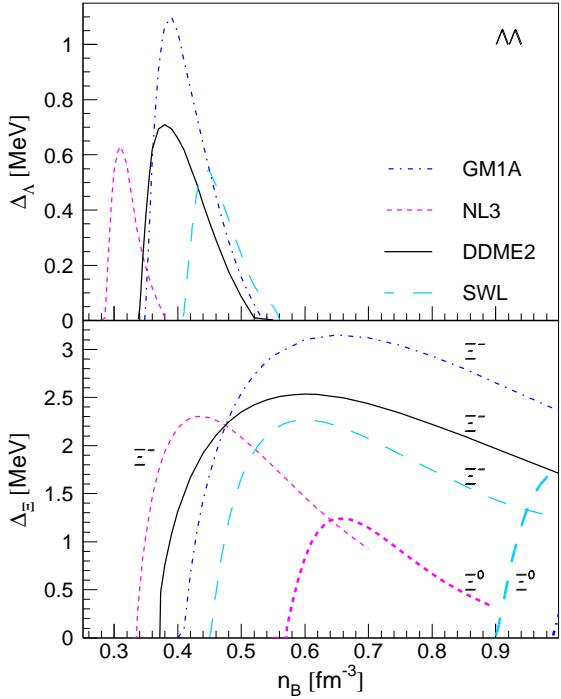}
\end{center}
\caption{The same as in Fig. \ref{fig:Gaps_kF}, but for $^1S_0$
  pairing gaps as a function of baryonic number density.  Note that
  the $\Xi^{0}$ hyperon appears in the density range shown in the
  figure only for the NL3 and SWL DFs and its onset for GM1A DF is at 
  $n_B = 0.99$ fm$^{-3}$, i.e., close to the upper limit of the range considered. }
\label{fig:Gaps_nB}
\end{figure}

The dependence of the pairing gaps on the composition of matter is displayed 
in Fig.~\ref{fig:Gaps_nB}.  
Note that the density range of $\Lambda\Lambda$ pairing is restricted to densities $n_B\le 0.55$
fm$^{-3}$ which implies that at high densities, which may be achieved in massive
stars, regions of unpaired $\Lambda$ matter will exist. In contrast to this, the
$\Xi^{-}$ component remains paired up to the highest densities. It can also be
seen that the NL3 model predicts density ranges for $\Lambda$, $\Xi^-$ and $\Xi^0$ pairing,
which deviate strongly from other, better constrained models. 
The largest discrepancy appears for the $\Xi^0$ pairing range, 
which is due to the early onset of $\Xi^0$s for this model.

\section{Neutrino radiation processes}
\label{sec:nu_processes}

The neutrino radiation from compact star interiors depends sensitively
on the particle content of matter and the magnitude and density
dependence of the pairing gaps of fermions. The leading neutrino
radiation processes in nucleonic phases are well-known [see for 
reviews~\cite{weber_book,Yakovlev_2001,Page_2004,2007PrPNP..58..168S}]. The
presence of hyperons leads to the hyperonic dUrca
processes~\citep{1992ApJ...390L..77P}, which can be written
symbolically as
\begin{eqnarray} 
\label{eq:UrcaLambda}
\Lambda &\to& p + l  + \bar\nu_l,\\
\label{eq:UrcaSigmaminus}
\Sigma^- &\to& \left(\begin{array}{c} n  \\
                       \Lambda \\
\Sigma^0
\end{array} \right) + l + \bar\nu_l,\\
\label{eq:UrcaXiminus}
\Xi^- &\to& \left(\begin{array}{c} \Lambda  \\ 
                                                         \Xi^0
                        \\
\label{eq:UrcaSigmazero}
\Sigma^0   \end{array} \right) + l + \bar\nu_l,\\
\label{eq:Xizero}
\Xi^0 &\to& \Sigma^+ + l  + \bar\nu_l,
\end{eqnarray}
where $l$ stands for a lepton, either electron or muon, and
$\bar\nu_l$ is the associated anti-neutrino.  Provided that all
hyperonic species involved in a given reaction exist in matter, the
corresponding thresholds on their density fractions are quite low - of
the order of a few percent. The hyperon abundances increase strongly as
soon as they become energetically favourable, therefore the hyperonic
dUrca processes start to operate soon after the onset of
hyperons. Their rates are larger than those of the modified Urca
processes involving hyperons~\citep{1987ApJ...316..691M} [which can be
visualized by adding a bystander baryon to the processes listed in
Eq.~\eqref{eq:UrcaLambda}-\eqref{eq:Xizero}].  Although, when allowed
by the triangle inequalities~\citep{Boguta1981,Lattimer1991}, the rate
of the nucleonic direct Urca process $n\to p + e + \bar\nu$ is higher
than its hyperonic counterparts, it is suppressed by the superfluidity
of nucleons.  Thus, hyperonic dUrca processes dominate if hyperons are
not paired.
\begin{figure}[t]
\begin{center}
\includegraphics[angle=0, width=0.99\columnwidth]{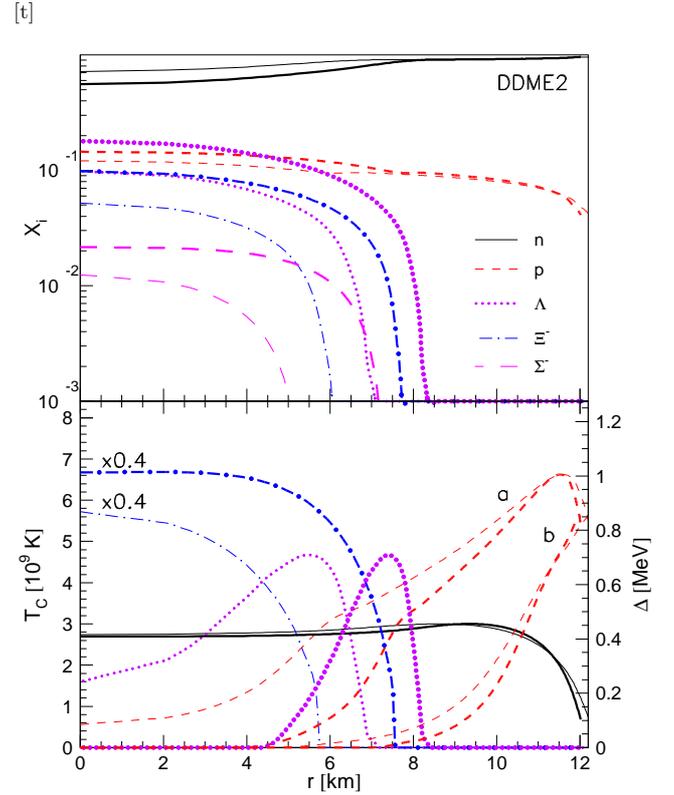}
\end{center}
\caption{ Upper panel: composition of NS with masses of 1.8$M_{\odot}$
  (thin) and 2$M_{\odot}$ (thick lines) in terms of $n$, $p$,
  $\Lambda$, $\Xi^-$, and $\Sigma^-$ relative abundances, as predicted
  by the DDME2 \protect\cite{Fortin_PRC2016} EoS.  Lower
  panel: the pairing critical temperatures for $n$, $p$, $\Lambda$,
  and $\Xi^-$ as a function of radial distance from the centre of the
  star.  Two scenarios for proton $^1S_0$ pairing have been
  considered: (a) CCDK \protect\cite{Chen_NPA1993} and (b) BCLL
  \protect\cite{Baldo_NPA1992}.  The relation 
$T_c [10^{10} K] \approx 0.66 \Delta [MeV]$
  has been used to find $T_c$ for the $^1S_0$ and $^3P_2-^3F_2$
  pairing gaps.  The pairing gaps for $\Xi^-$s are scaled by a factor
  of 0.4.  }
\label{fig:compoDDME2}
\end{figure}
\begin{figure}[t]
\begin{center}
\includegraphics[angle=0, width=0.99\columnwidth]{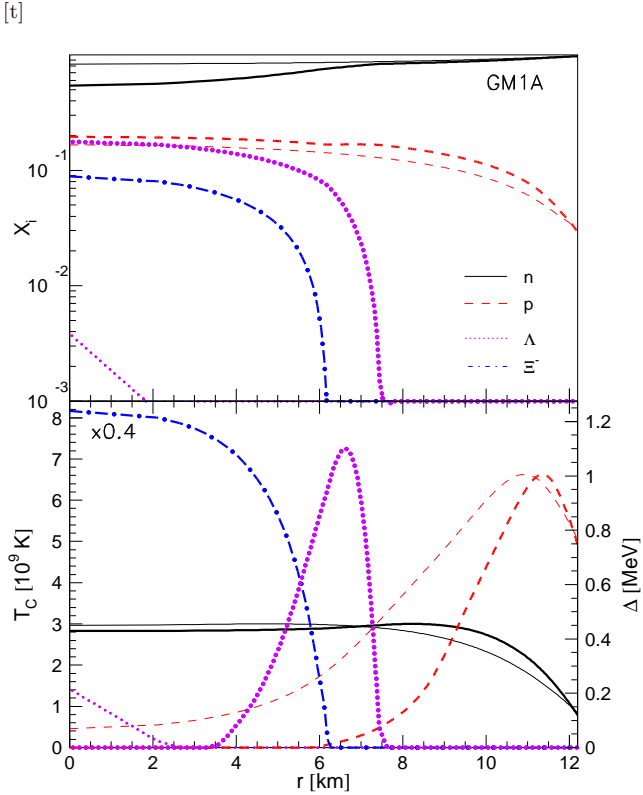}
\end{center}
\caption{The same as Fig. \ref{fig:compoDDME2},
  but for NS with masses 1.5$M_{\odot}$ (thin) and 1.9$M_{\odot}$
  (thick lines) and the GM1A model \protect\citep{Gusakov_MNRAS2014}. 
  The pairing gaps for $\Xi^-$s are scaled by a factor of 0.4.  
  The proton $^1S_0$ pairing gap is implemented as in
  \protect\cite{Chen_NPA1993} and the neutron pairing gap is chosen as in
  Fig. \ref{fig:compoDDME2}.  }
\label{fig:compoGM}
\end{figure}
Hyperonic pairing, discussed in the previous section, changes the
picture in two-fold way: first, it suppresses the hyperonic dUrca rate
(exponentially at low temperatures $T\ll T_{cY}$, where $T_{cY}$ is
the critical temperature of hyperonic pairing); secondly, it opens a
new channel of neutrino emission based on the Cooper pair-breaking and
formation (PBF) mechanism: 
\begin{eqnarray}
\label{eq:Y_PBF}
\{YY\} \to Y+Y + \nu + \bar \nu, \quad 
Y+Y\to \{YY\} + \nu + \bar \nu, 
\end{eqnarray}
where $\{YY\}$ stand for a hyperonic Cooper pair. The rate of the
process \eqref{eq:Y_PBF} has been discussed
previously~\citep{1999A&A...343..650Y,2001PhLB..516..345J}, but it
requires some revision. Specifically, the component of the emissivity
to the vector current coupling ($\propto c_V^2$) is negligible for
$S$-wave paired baryons compared to that of axial-vector coupling
($\propto c_A^2$)~\citep{2006PhLB..638..114L,2007PhRvC..76e5805S,
  2008PhRvC..77f5808K,2010PhRvC..81f5801K,2009PhRvC..79a5802S,2012PhRvC..86b5803S}.
The first contribution scales as $(v_{YF}/c)^4$, where $v_{YF}$ is the
Fermi velocity of a given hyperon and $c$ is speed of light, whereas
the second one scales as $(v_{YF}/c)^2$. This last contribution can be
written, in analogy with the result for nucleons, as
\begin{eqnarray}
\epsilon_{Y} &=& \frac{4G_F^2\zeta_A}{15\pi^3\hbar^{10}c^6} [c_{A}^2\nu_0
v_{F}^2  I_{\nu}]_Y ~T^7, 
\end{eqnarray}
where the phase-space integral is given by 
\begin{eqnarray}
I_{\nu Y} &=& z_Y^7 \int_1^{\infty} dy \frac{y^5}{\sqrt{y^2-1}} f_F \left(z_Y y\right)^2 ,
\end{eqnarray}
where $G_F$ is the Fermi coupling constant, $\zeta_A = 6/7$, the
notation $[\dots]_Y$ indicates that the quantities in the braces
depend on the hyperon type $Y$, $c_A$ is the axial-vector coupling
constant, $\nu_0 = m_L^* p_{F}/\pi^2$ is the density of states,
$f_F (x)= [\exp(x) + 1]^{-1}$ is the Fermi distribution function with
$z = \Delta/T$ and $m_L^*$ refers to the Landau effective mass (as
opposed to the Dirac effective mass entering in the defintion of
eigenstates of the Dirac equation for nucleons and hyperons in
medium).  Note that we have used the Landau effective masses in
expressions for the emissivities of the weak processes.  The axial-vector
current coupling constants for the tree-level $Y\to Y$ transitions are
given by~\cite{1997PhRvD..55.5376S}
\begin{eqnarray}
&&c_{A}(\Lambda) = - (F+D/3) = -0.73,\\
&&c_{A}(\Xi^-) = c_{A}(\Sigma^-) =  D-3F = -0.58,\\
&&c_{A}(\Xi^0) = -c_{A}(\Sigma^+) =  -( D+F) = -1.26,\\
&&c_{A}(\Sigma^0) = D-F = 0.34,
\end{eqnarray}
where the parameter values used in the numerical evaluation are
$D = 0.79$ and $F = 0.47$.  Finally we note that the hyperons
contribute also to the specific heat of the core of the star; these
contributions are suppressed once they pair to form a condensate. The
heat capacity and its suppression by $S$-wave superfluidity are
modelled in full analogy to the $S$-wave paired nucleons.

To understand the relative role of the hyperons played in the cooling
of NS, it is useful first to examine the relative
abundances of the baryon octet $x_i = n_i/n$, where $n_i$ is the
partial density of the baryons, and the critical temperatures $T_{ci}$
of their pairing phase transition as a function of some interior
parameter, for example, the internal radius.
Figs.~\ref{fig:compoDDME2} and \ref{fig:compoGM} show these
dependences for the DDME2 and GM1A models; the results for the SWL
model are very close  to that of DDME2 and are not shown. 

The profiles of the relative abundances reveal that aside from
dominant component of neutrons with $x_n\le 1$ the baryons separate
into two groups: in the first group, which includes $p$, $\Lambda$ and
$\Xi^-$ the baryon abundances are of the order of 0.1.  The second group
includes $\Sigma^-$ and (at high densities) $\Xi^0$ and, possibly, other hyperons
with relative abundances $\le 0.01$. This latter group plays a
negligible role in the cooling, except of $\Sigma^-$ which do not pair
under our working assumption that the relevant interaction is
repulsive. Among the second group the first baryons to pair are
$\Xi^-$s. Their maximal critical temperature $\sim 2\times 10^{10}$ K;
a burst of neutrino emission via PBF processes from this component is
overshadowed by the Urca process on $\Lambda$ hyperons given by
Eq.~\eqref{eq:UrcaLambda}.  The remaining baryons $n$, $p$ and
$\Lambda$ thus control cooling through the Urca processes, whereby
the following factors play a role: (a) whether or not the nucleonic
dUrca threshold is achieved in the star interior; (b) whether some of
these baryons lose their pairing at high density leaving some interior
regions unpaired. As seen, e. g., from Fig.~\ref{fig:compoDDME2} the
more massive star's interior is stripped from $\Lambda$ and $p$
pairings, which provides rapid Urca cooling via the $(\Lambda p)$
channel (see Eq.~\eqref{eq:UrcaLambda}). Note that the $\Xi^-$ hyperons
are paired in the entire range of their existence and cannot
contribute via the dUrca process (except the special case where only
the tale of the pairing gap enters the density range of the star).

Thus, we conclude that the main effect of accelerated cooling is
caused by absence of pairing in $p$ and $\Lambda$ components in the
high density regions of the stars: the more massive is the star, the
larger is the relevant region. Comparing the two models of dense
matter, i.e., Figs.~\ref{fig:compoDDME2} and \ref{fig:compoGM} it is
easy to conclude that the same arguments apply also in the case of
matter composition based on GM1A model. The minor differences (for
example, the absence of $\Sigma^-$ hyperons, or larger fraction of
protons) do not affect the mechanism by which the rapid cooling
becomes available with increasing stellar mass.

\section{Thermal evolution of hypernuclear stars}
\label{sec:therm_ev}

\begin{figure}
\begin{center}
\includegraphics[angle=0, width=0.99\columnwidth]{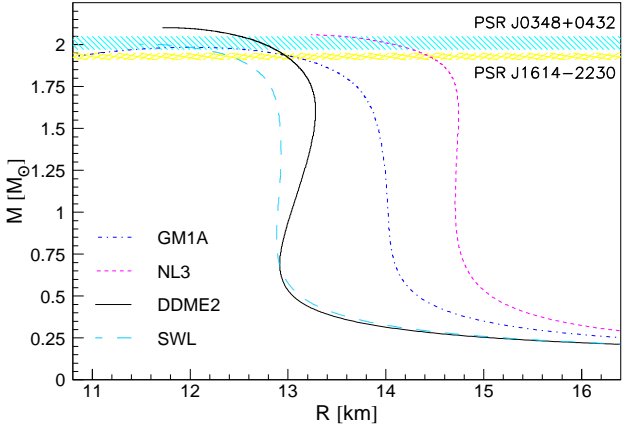}
\end{center}
\caption{Mass versus radius relation for the stellar models considered in this work. 
  The shaded areas show the masses of two massive pulsars, PSR J0348+0432 with
  $M=2.01\pm 0.04 M_{\odot}$ \protect\citep{2013Sci...340..448A}, 
  and PSR J1614-2230 with $M=1.93\pm 0.02
  M_{\odot}$ \protect\citep{Fonseca2016}. }
\label{fig:MR}
\end{figure}

The above discussed models of the EoS of hypernuclear matter, which
are based on covariant relativistic DF theory, were
employed to construct static, spherically symmetrical configurations of
self-gravitating objects assuming that these are non-rotating and
nonmagnetized. Fig.~\ref{fig:MR} shows the mass versus radius relation
for our collection of EoS, in particular it demonstrates that our
models satisfy the astrophysical constraints placed by pulsar mass
measurements.

These models have been further evolved from some initial temperature
distribution (chosen to be sufficiently large, but the details are
inessential) at initial time assuming that the structure of the models
does not change in time. The evolution was followed for $10^6$ yr after
which the star surface temperature drops below the observable
limit. We employ the public domain {\sc NSCool} code
\footnote{www.astroscu.unam.mx/neutrones/NSCool} by D. Page, which was suitably
modified to include the physics of hyperonic components.  In all
considered cases heating sources, magnetic fields and accretion have
been disregarded.  The envelope is assumed to consist of Fe.  The
crust EoS of \cite{NEGELE1973298} and \cite{1989A&A...222..353H} was
smoothly merged with the core EoS.  Finally the following computations
of the nucleonic pairing gaps are implemented in the code. The neutron
$^1S_0$ pairing in the crust is given by \cite{SFB_2003}.  For neutron
$^3P_2-^3F_2$ pairing we choose the 'b' curve of the gap shown in
fig. 10 of \cite{Page_2004}. Two computations of $^1S_0$ proton
pairing have been used: ``CCDK''  by \cite{Chen_NPA1993} and ``BCLL''
by \cite{Baldo_NPA1992}. These computations differ mostly in the density
domain that proton superfluid occupies: $0 \lesssim k_F \lesssim 1.3$
fm$^{-1}$ for CCDK and $0.1 \lesssim k_F \lesssim 1.05$ fm$^{-1}$ for
BCLL. The difference in the maximal values of the critical
temperatures for these models is about 20$\%$,
$T_c\approx 6.6 \times 10^9$ K for CCDK and $\approx 5.6 \times 10^9$~K
BCLL.
\begin{figure}
\begin{center}
\includegraphics[angle=0, width=0.99\columnwidth]{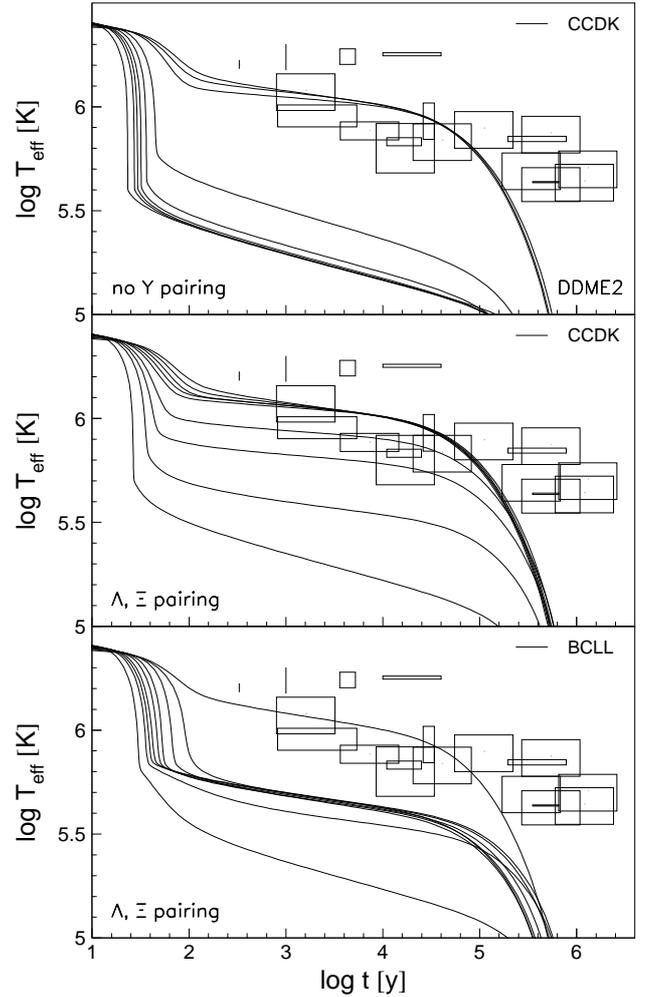}
\end{center}
\caption{Cooling models based on the DDME2 EoS for NS masses 1.3, 1.4,
  1.5, 1.6, 1.7, 1.8, 1.85, 1.9 and 2$M_{\odot}$ (from top to bottom)
  without (top panel) and with (middle and bottom panels) hyperon
  pairing.  Two scenarios for $^1S_0$ proton pairing have been used:
  CCDK \protect\citep{Chen_NPA1993} (top and middle panels) and BCLL
  \protect\citep{Baldo_NPA1992} (lower panel).  Observational data
  correspond to 19 isolated NS listed in \protect\cite{Beznogov_2015}.
}
\label{fig:Teff_DDME2}
\end{figure}

\subsection{Cooling models without nucleonic Urca process}

We first discuss models where the process $n\to p + e + \bar\nu_e$ and
its inverse are forbidden by sufficiently low proton fraction.

We start with the cooling models based on the DDME2 EoS and show the
dependence of the effective surface temperature $T_{\rm eff}$ on time
in Fig.~\ref{fig:Teff_DDME2} for the cases of unpaired (upper panel)
and paired (middle panel) hyperon component for the proton pairing
model CCDK~\citep{Chen_NPA1993}. The alternative proton pairing
model BCLL~\citep{Baldo_NPA1992} with paired hyperon component is
shown in the lower panel of the same figure.  The data shown in the
figure (including the error bars) are the temperatures inferred from the
thermal component of the X-ray emission measured from a number of
pulsars; the pulsar ages are the spin-down ages unless there is an
association with a known supernova [see \cite{Beznogov_2015}].

In the absence of hyperonic pairing the cooling curves for different
masses separate into sets corresponding to slowly cooling stars with
surface temperatures $T\sim 10^6$~K and fast cooling stars with
surface temperatures by an order of magnitude lower at about
$t\sim 10^{4}$ yr. This dichotomy can be understood by examining the
neutrino luminosity of neutrino processes. For the DDME2 model the
threshold for the nucleonic dUrca process is not reached for any
model, therefore the only dUrca processes available are those on
hyperons. For star masses $M/M_{\odot} \simeq 1.5$ the dominant
hyperonic process is the dUrca process on $\Lambda$s
\eqref{eq:UrcaLambda} but it is effective only at early stages of
evolution (e.g., for $M/M_{\odot} = 1.5$ for log\,$t\le 3.2$~yr) and
is suppressed after protons become superfluid.  For
$M/M_{\odot} = 1.6$ model the central density exceeds the $\Xi^-$
threshold, as a consequence the purely hyperonic Urca process
\eqref{eq:UrcaXiminus} in the $(\Xi^-\Lambda)$ channel dominates.  For
$M/M_{\odot} > 1.7$ the central density exceeds the $\Sigma^-$
threshold and another purely hyperonic Urca process
\eqref{eq:UrcaSigmaminus} in the $(\Sigma^-\Lambda)$ channel becomes
the dominant agent [the process \eqref{eq:UrcaXiminus} involving
$\Xi^-$ is about $10\%$ of the neutrino luminosity].  Note that the
$\Sigma^-$ decay into neutrons is suppressed by neutron pairing and
does not play a role.  In the most massive model $M/M_{\odot} = 2.0$
the process $(\Lambda p)$ channel \eqref{eq:UrcaLambda} contributes
again, because for such massive models the proton pairing vanishes in
the inner core ($r\le 4$~km).
     
In the case where the hyperon pairing is included, models fill-in the
region between the two extremes discussed above. For stars with
$M/M_{\odot} \ge 1.7$, the dUrca process in the $(\Lambda p)$ channel
\eqref{eq:UrcaLambda} dominates the early evolution for
log\,$t\le 4$~yr and is suppressed at the later times by pairing. The
process \eqref{eq:UrcaXiminus} in the $(\Xi^-\Lambda)$ channel does
not contribute significantly in any range of masses, because of strong
suppression by their superfluidity. For stars with
$M/M_{\odot} \ge 1.8$ the hyperonic dUrca process
\eqref{eq:UrcaSigmaminus} involving $\Sigma^-$s (which do not pair) is
the dominant process up to times log\,$t\le 5$ to $5.5$~yr, after
which the photon cooling from the surface takes over. For the massive
model with $M/M_{\odot} = 2.0$ the inner core ($r\le 4$~km) features
not only unpaired protons, but also unpaired $\Lambda$s. Nevertheless
the process \eqref{eq:UrcaSigmaminus} dominates, i.e, {\it for massive
  hyperonic compact stars the dominant cooling is provided by purely
  hyperonic dUrca process.}

The cooling models are sensitive to the proton pairing pattern inside
the star because the dUrca process in the $(\Lambda p)$ channel
\eqref{eq:UrcaLambda} is the most effective process in low-mass
stars. In particular, whenever the proton pairing gaps are small and
regions exists where protons are unpaired (see
Fig.~\ref{fig:compoDDME2}) this process dominates the cooling. This is
indeed the case for BCLL pairing, as shown in the lower panel of
Fig.~\ref{fig:Teff_DDME2}. Apart from the lightest star, the remaining
configurations are effectively cooled by this process, which leads to
clustering of cooling curves in the fast cooling regime. In the
heaviest stars $M/M_{\odot} \ge 1.9$ the competing process on
\eqref{eq:UrcaSigmaminus} in the $(\Sigma^-\Lambda)$ channel dominates.
In this case the thermal evolution is the same in both models of
protonic pairing (c. f. the middle and lower panels of
Fig.~\ref{fig:compoDDME2}).
\begin{figure}
\begin{center}
\includegraphics[angle=0, width=0.99\columnwidth]{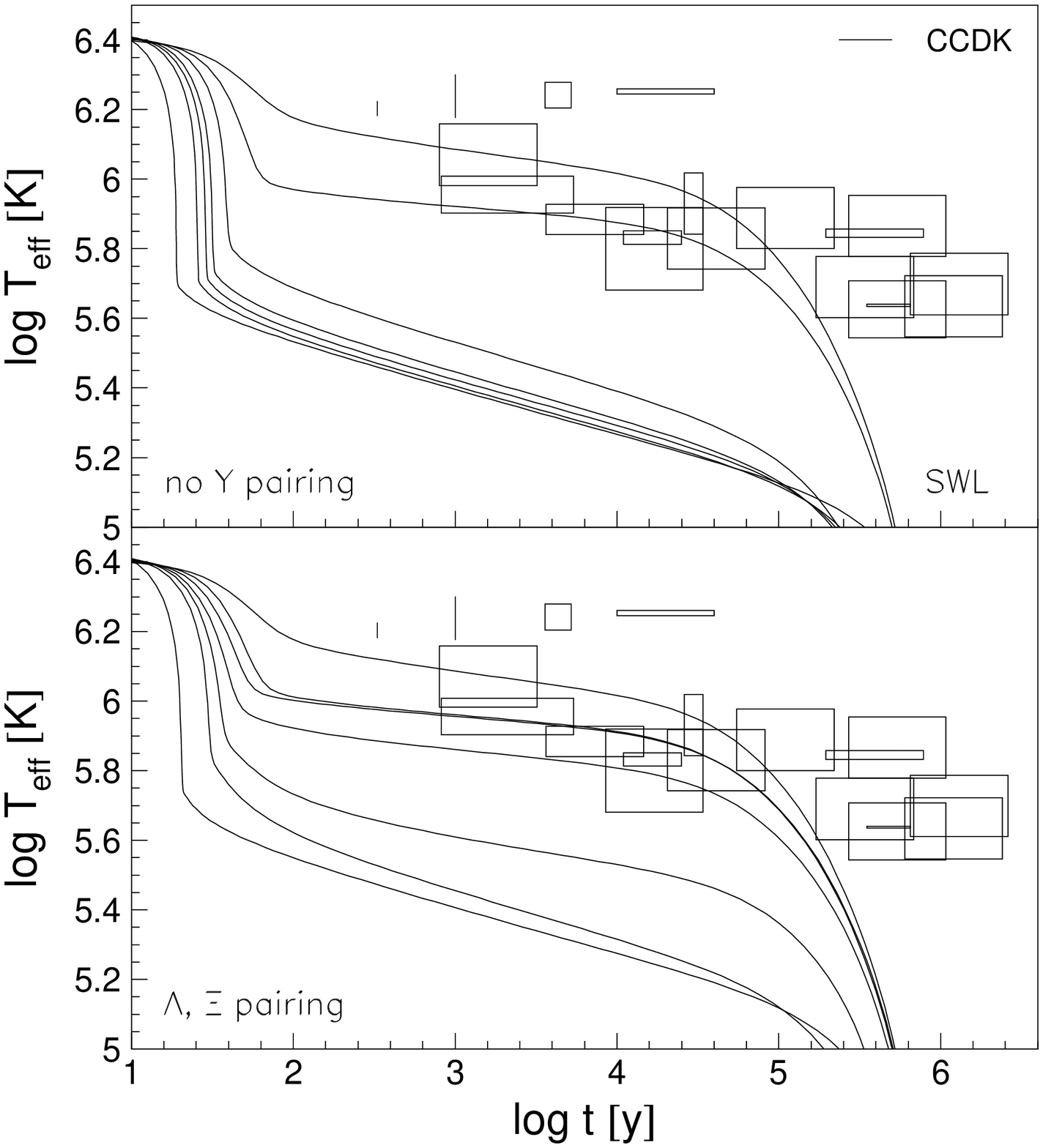}
\end{center}
\caption{Cooling models based on the  SWL EoS for NS masses
  1.5, 1.6, 1.7, 1.8, 1.85, 1.9, 2.0$M_{\odot}$ (from top to bottom).  For
  $^1S_0$ proton pairing the  CCDK \protect\citep{Chen_NPA1993} gap has been used.
 }
\label{fig:Teff_SWL}
\end{figure}

Next we consider cooling models based on the SWL equations of state,
which are shown in Fig.~\ref{fig:Teff_SWL}.  The difference to the
previous case is the absence of unpaired $\Sigma^-$ hyperons,
therefore this model illustrates the physics of cooling of compact
stars where {\it all hyperons pair.}  As in the previous case the
cooling curves separate into slow and fast cooling sets if hyperon
pairing is ignored. The slow cooling set contains two models with
$M/M_{\odot}= 1.5,\, 1.6$ which are either purely nucleonic or contain
only a small admixture of $\Lambda$s. The increase of $\Lambda$
abundances and the onset of other hyperons accelerates the cooling and
the models with $M/M_{\odot} = 1.7$ to 2 form the second set of
rapidly cooling stars. The $M/M_{\odot} = 1.7$ and 1.8 models cool
predominantly via the $(\Xi^-\Lambda)$ channel of
Eq.~\eqref{eq:UrcaXiminus}, because proton pairing suppresses the
former process. For models $M/M_{\odot} \ge 1.9$ the proton pairing
vanishes in the central region ($r\le 4$~km for 1.9$ M_{\odot}$ star)
of the star and both processes \eqref{eq:UrcaLambda} and
\eqref{eq:UrcaXiminus} contribute nearly equally. 

Including the hyperon superfluidity completely suppresses the 
process \eqref{eq:UrcaXiminus} on $\Xi^-$ particles, which have quite
large gap.  The remaining dUrca process \eqref{eq:UrcaLambda} in the
$(\Lambda p)$ channel dominates the neutrino luminosity up to
log$t\,\le 4.5$ for models $M/M_{\odot} \ge 1.7$ and 1.8, but its rate
is suppressed by the $\Lambda$ pairing, consequently the cooling
tracks pass through the area where the cooler set of the observed
stars is located. The models with $M/M_{\odot} \ge 1.9$ are not
affected by hyperonic pairing, because, as pointed out above, these
develop a core where $\Lambda$ and proton pairing vanish. Thus, we
conclude that in the case where $\Sigma^-$ does not appear in matter
{\it and} cooling via processes involving $\Xi^-$ are strongly
suppressed by their superfluidity the dominant role is played by the
dUrca process on $\Lambda$s, which provides a good description of the
data of the observed cooler compact stars. However, when the cores of
most massive stars develop unpaired regions with $\Lambda$ hyperons
and protons the cooling is strongly accelerated and the surface
temperatures drop well below the observed ones.

\subsection{Allowing for nucleonic direct Urca process}

\begin{figure}
\begin{center}
\includegraphics[angle=0, width=0.99\columnwidth]{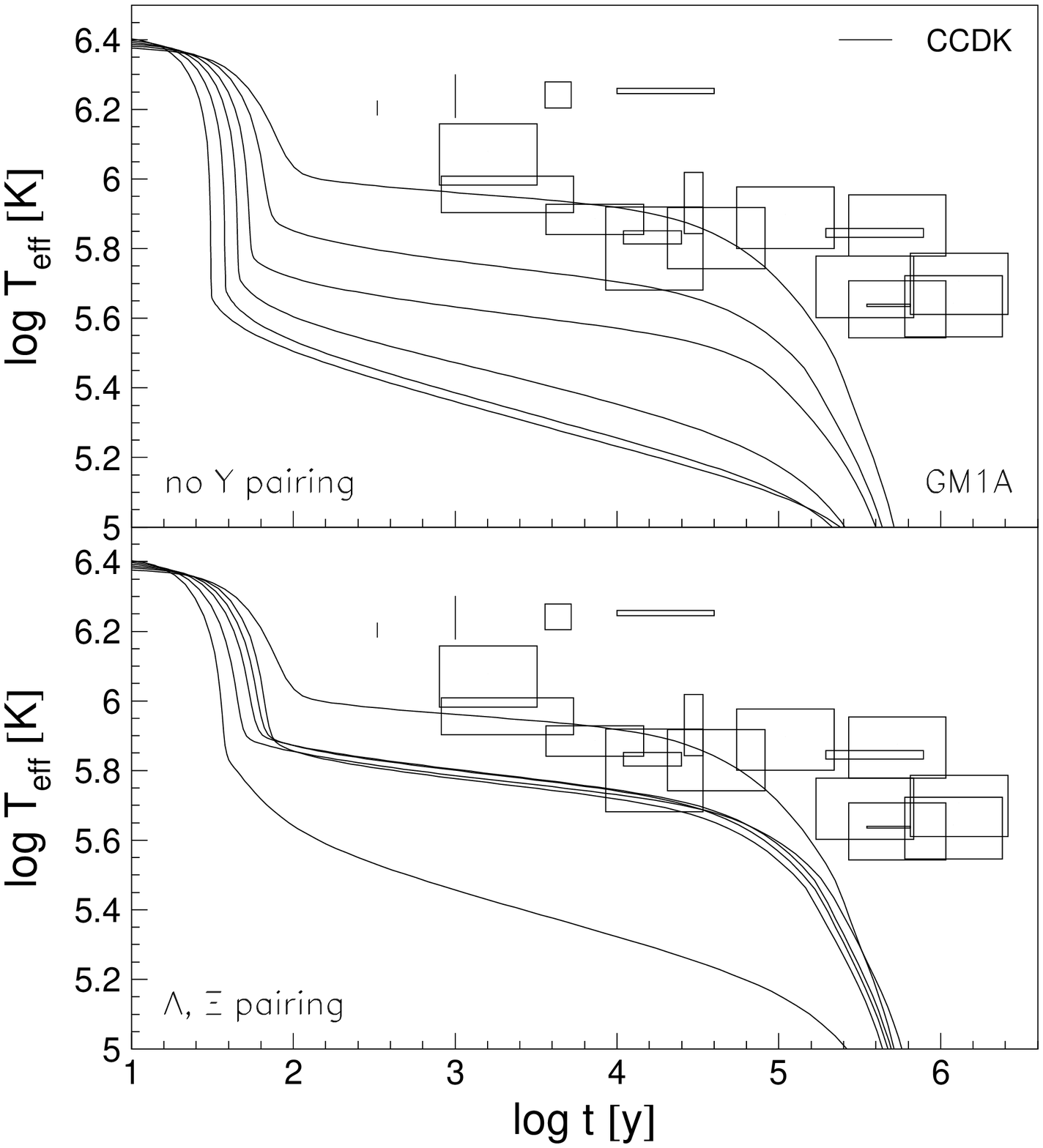}
\end{center}
\caption{Cooling models based on the GM1A \protect\citep{Gusakov_MNRAS2014}
  EoS for NS masses 1.4, 1.5, 1.6, 1.7, 1.8, $1.9M_{\odot}$ (from top
  to bottom).  For $^1S_0$ proton pairing the CCDK
  \protect\citep{Chen_NPA1993} gap has been used. }
\label{fig:Teff_GM1A}
\end{figure}

Now we turn to the GM1A models, which support the nucleonic dUrca
process and consider first the case of unpaired hyperons.  The purely
nucleonic model with $M/M_{\odot} = 1.4$ in this case cools
predominantly via the nucleonic dUrca process. For heavier models,
$M/M_{\odot} =1.5- 1.6$, the dominant neutrino radiation mechanism
becomes the hyperonic dUrca process in the $(\Lambda p)$ channel, as
the nucleonic dUrca process is suppressed by the neutron and proton
gaps.  For the model with $M/M_{\odot} \simeq 1.7$ the cooling is
controlled equally by the processes \eqref{eq:UrcaXiminus}
$(\Xi^-\Lambda)$ channel and \eqref{eq:UrcaLambda} by $(\Lambda p)$
channel up to time-scales log$\,t\le 3$~yr after which the last process
is suppressed by proton pairing, whereas the first one operates at
full strength. Nucleonic dUrca does not play any substantial role in
these models during the neutrino cooling era, as it is effectively
suppressed by nucleonic pairing. For models with $M/M_{\odot} \ge 1.8$
proton pairing vanishes in the inner core ($r\le 4$~km for
$M/M_{\odot} =1.7$ and $r\le 6$~km for $M/M_{\odot} =1.9$).  As
a consequence the hyperonic dUrca process \eqref{eq:UrcaLambda} on
unpaired $\Lambda$s operates at full strength in the $(\Lambda p)$
channel. Even though protons are unpaired, the neutron pairing
suppresses the nucleonic Urca in the neutrino cooling era and it again
is unimportant.  As seen in Fig.~\ref{fig:Teff_GM1A}, upper panel, the
emergence of hyperons leads to lower temperatures of compact stars;
for masses in the range below 1.6 $M/M_{\odot}$, which is due to dUrca
process \eqref{eq:UrcaLambda} on $\Lambda$s, a stronger shift towards
lower temperatures occurs for $M/M_{\odot} \geq 1.7$ models due to the
onset of the dUrca process \eqref{eq:UrcaXiminus} on $\Xi^-$.  The
lowest effective temperatures of models with $M/M_{\odot}=1.8$ and 1.9 are
explained by the fact that, due to the high densities reached in the
core, the proton pairing gap in the $^1S_0$ channel vanishes and, thus,
the process \eqref{eq:UrcaLambda} operates at full strength over a
large fraction of the core.

In the case where hyperon superfluidity is included, the situation is
simpler. Because of the large pairing gap of $\Xi^-$ the dUrca
processes \eqref{eq:UrcaXiminus} on $\Xi^-$ do not play any role. The
nucleonic Urca is suppressed still as in the previous case.  The
remainder dominant neutrino emission process is dUrca in the
$\Lambda p$ channel \eqref{eq:UrcaLambda}.  As a consequence, the
stars with masses up to 1.8 $M/M_{\odot}$ remain relatively warm
through thermal evolution, with their tracks clustered at the lower
edge of the observed NS temperatures. The sharp drop in the
temperature observed for the model with 1.9 $M/M_{\odot}$ is due to
the unpairing of $\Lambda$s at high densities, i.e., the closing of
their $^1S_0$ gap. This occurs in the density range where proton
$^1S_0$ gap closes as well. In the absence of pairing the dUrca
process in the $(\Lambda p)$ channel operates at full strength leading
to minimal possible surface temperatures of the most massive models as
discussed above.

\section{Conclusions}
\label{sec:conclusions}

In this work we considered a number of models of hypernuclear matter
based on the covariant DF theory which are compatible with the
two-solar-mass constraint \citep{2013Sci...340..448A,Fonseca2016}
on dense matter and semi-empirical depths of potential
wells of hyperons in symmetric nuclear matter at saturation density
\citep{Millener1988}. 
We have solved the BCS equations in the hyperonic sector and obtained 
the gaps of hyperons in the $\Lambda$ channel as well as, for the first time, 
in the $\Xi^-$ and $\Xi^0$ channels. With this input, we carried out a series of
cooling simulations of compact stars and compared them with the
available data.

Our study of the cooling of hypernuclear compact stars reveals that
quite generally the hyperonic component plays a dominant role in
neutrino cooling of hypernuclear stars, even in the case where the
nucleonic dUrca is allowed by the composition of matter. The main
cooling agents are various flavours of the hyperonic dUrca process
listed in equations~\eqref{eq:UrcaLambda}-\eqref{eq:Xizero}.  We have
included in our studies the pair-breaking processes on
hyperons~\eqref{eq:Y_PBF}, but these turned out to be subdominant to
the dUrca processes.  The obtained cooling behaviour of our models,
shown in Figs.~\ref{fig:Teff_DDME2}--\ref{fig:Teff_GM1A}, depends
sensitively on the details of the composition of matter they predict
and pairing gaps. In the following we summarize the general trends
found from these simulations.

\begin{itemize} 

\item {\it Mass-hierarchy.}  Consider a sequence of compact
  hypernuclear stars arranged from the lightest to the heaviest
  ones. We now follow the changes in their cooling behaviour along
  such a mass hierarchy.  Hyperons start to populate the interiors of
  compact stars with masses $M/M_{\odot}> 1.5$; in all models the
  first hyperons to appear is $\Lambda$ and the corresponding dUrca
  process on ($\Lambda, p$) plus leptons (see
  Eq.~\eqref{eq:UrcaLambda}) is the dominant cooling agent in the
  neutrino cooling era, even in the case where nucleonic dUrca is
  kinematically allowed. The efficacy of the dUrca process on
  ($\Lambda, p$) is the result of moderate critical temperature of
  condensation of $\Lambda$s, with a maximum in the range
  $T_{c,\rm max}\simeq 5-7\times 10^9$K. For stars with
  $M/M_{\odot}> 1.6$ our models predict the onset of $\Xi^-$, which
  would have provided the dominant cooling mechanism via
  ($\Xi^-\Lambda$) channel of the dUrca process \eqref{eq:UrcaXiminus}
  in the absence of hyperonic pairing. However the large critical
  temperatures of pairing $T_{c\Xi} \simeq 2\times 10^{10}$K prevent $\Xi^-$s
  from playing a role in cooling of the star.  For stars with
  $M/M_{\odot} \ge 1.6$ $\Sigma^-$ appear in the DDME2 model, but not
  in others models.  Their appearance implies that the dominant
  neutrino cooling process is the ($\Sigma^-\Lambda$) channel of the
  dUrca process \eqref{eq:UrcaSigmaminus}, which accelerates the
  cooling rate.  Finally, a new feature in most massive stars
  $M/M_{\odot}\simeq 2$ is the unpairing of $\Lambda$s {\it and}
  protons in the central core of the star, because their large density
  renders $^1S_0$ pairing interaction repulsive. Then, again, the
  dUrca process in the ($\Lambda, p$) channel \eqref{eq:UrcaLambda}
  dominates the cooling, but at a larger rate characteristic for
  unpaired hyperonic matter.

\item {\it Hyperonic species.} The $\Lambda$ hyperon appears
  first and plays a substantial role when the densities are low enough
  that the $\Xi^-$ does not nucleate and when the density is so
  high that they do not form pairs in the $^1S_0$. The main neutrino
  emission channel is the  dUrca process \eqref{eq:UrcaLambda} in the
  ($\Lambda, p$) channel.
  
  The next hyperon to appear is $\Xi^-$. It does not play a role in
  the cooling, because of its large condensation temperature
  $T_{c\Xi}\simeq 2\times 10^{10}$\,K and wide pairing gap.  An exception can
  arise in a narrow mass range around $M/M_{\odot}\simeq 1.6$, where
  the average density of $\Xi^-$-gas is low, therefore only the
  low-density 'tail' of the pairing gap function is important.

  The $\Sigma^-$ hyperon nucleates in one out of three models
  considered here within sufficiently massive (but stable)
  hypernuclear stars. If $\Sigma^-$ do not pair, as implied by a repulsive 
  $\Sigma^-$ interaction, $\Sigma^-$ contribute to the
  dominant cooling mechanism via the ($\Sigma^-\Lambda$) channel in
  Eq.~\eqref{eq:UrcaSigmaminus}. Large neutron pairing throughout the
  hypernuclear core does not allow for the ($\Sigma^- n$) channel to
  operate.

  The fractions of other hyperons, in particular $\Xi^0$, never become
  significantly large in our models to be important for the neutrino
  cooling.

 \item {\it Consistency with the data.}  The observational data
   requires a set of cooling tracks covering the range of
   temperatures $ 5.7\le \log T_{\rm eff}[{\rm K}]\le 6.3$ in the
   neutrino cooling era $\log t\le 5\,$[yr]; the required variations
   of cooling tracks can be achieved by varying the masses of the
   object along the sequence defined by an EoS. We find that the
   DDME2 model can account for this, with the lightest stars
   (featuring hyperonic cores) $M/M_{\odot}\le 1.6$ accounting for
   hotter objects and the more massive ones $M/M_{\odot}\le1.85$
   accounting for the cooler objects (see
   Fig.~\ref{fig:Teff_DDME2}). An important ingredient of this picture
   is the proton $^1S_0$ gap (CCDK model) extending to large
   densities; if the proton $^1S_0$ pairing gap is narrow (as exemplified by the BCLL
   model), then the hypernuclear stars cool too fast. The SWL model
   shows an analogous behaviour (see Fig.~\ref{fig:Teff_SWL}).  In the
   case of GM1A model, which allows for nucleonic dUrca process, the
   cooling of all the models except the most massive one is at the
   lower edge of the observable band of surface temperatures of
   thermally emitting NS. Because this model does not feature a
   $\Sigma^-$ hyperon, the sharp drop in the temperature for the most
   massive member shown in Fig.~\ref{fig:Teff_GM1A} is only due to the
   unpairing of $\Lambda$s and protons at high densities and efficacy
   of the dUrca process in the $(\Lambda p)$ channel.  

   Thus, we conclude that the hypernuclear models where $\Sigma^-$
   hyperon does not nucleate can account for surface temperatures of
   the cooler class of thermally emitting compact stars and,
   inversely, hypernuclear stars should be observable in soft X-rays
   through their thermal emission from the surface, unless they are
   extremely massive, i.e., $1.9\le M/M_{\odot}\le 2$ (see however
   below).

 \item {\it Alternatives.} We now discuss the physical
   alternatives to the key features discussed above. First, our models
   are based on the evidence of highly repulsive interaction between
   $\Sigma^-$ and nucleons, which has the consequence that the onset
   of $\Sigma^-$ is shifted to higher densities; we also assume a
   repulsive $\Sigma^-\Sigma^-$ interaction which allows us to
   neglect the $\Sigma^-$ pairing.  Should the interaction among
   $\Sigma^-$ and nucleons be less repulsive in dense matter $\Sigma^-$
   will replace $\Xi^-$s with no significant effect on the EoS. If in
   addition the $\Sigma^- \Sigma^-$ interactions are attractive
   \citep{Sasaki_2015}, $\Sigma^-$ pairing will lead to a suppression
   of the associated dUrca processes.  A study of the consequences of
   the possible interchange between the $\Sigma^-$ and $\Xi^-$ on NS
   thermal evolution is beyond the scope of the present paper and will
   be addressed elsewhere.  Secondly, the rapid cooling property of
   most massive models crucially depends on the closing of the $^1S_0$
   gaps for $\Lambda$s and protons at high densities. However,
   complete unpairing can be avoided if a higher partial wave channel,
   such $^3P_2-^3F_2$ coupled channel (which is known to be attractive
   in the case of nucleons) or even possibly an attractive $D$-wave
   channel, provides sufficient attraction to generate gaps and
   critical temperatures of relevant magnitude ($T_c\geq 10^8$K). In
   that case the extremely massive hypernuclear stars
   ($1.9\le M/M_{\odot}\le 2$) will undergo a slower cooling evolution
   than claimed above.

\end{itemize}

\section*{Acknowledgments}

AR thanks Constan\c{c}a Provid\^encia for discussions and acknowledges
the NewCompStar COST Action MP1304 for the partial support of this
project through an STSM grant as well as the hospitality of the
Frankfurt Institute for Advanced Studies.  AS is supported by the
Deutsche Forschungsgemeinschaft (Grant No. SE 1836/3-2).  FW is
supported by the National Science Foundation (USA) under Grants
PHY-1411708 and PHY-1714068.  We thank ECT$^*$ Trento for its generous
hospitality during the final stages of this work.

\bibliographystyle{mnras}

\end{document}